\documentclass[ amssymb, amsmath, showpacs, twocolumn, aps, prb, footinbib]{revtex4-1} 
\usepackage[normalem]{ulem}

\usepackage{graphicx,type1cm,eso-pic,color}

\renewcommand{\d}{\,\mathrm{d}}
\newcommand{\e}{\,\mathrm{e}} 
\newcommand{\imunit}{\mathrm{i}} 
\newcommand{\Tr}{\mathop{\mathrm{Tr}}} 
\newcommand{\tens}[1]{\overline{\overline{\mathsf{#1}}}} 
\newcommand{\Expint}[1]{\mathrm{E}_{#1}} 
\newcommand{\herm}{^\dagger}
\newcommand{\Refonlinecite}[1]{Ref.~\onlinecite{#1}}
\newcommand{\Refsonlinecite}[2]{Refs.~\onlinecite{#1} and \onlinecite{#2}}

\begin{document}
\title{Angular dependence, blackness and polarization effects in integral
 conversion electron M\"{o}ssbauer spectroscopy}

\author{Szil\'{a}rd~\surname{Sajti}} \email{sajti.szilard@wigner.mta.hu}
 \author{Ferenc~\surname{Tanczik\'{o}}}
 \author{L\'{a}szl\'{o}~\surname{De\'{a}k}}
 \author{D\'{e}nes~L.~\surname{Nagy}}
 \author{L\'{a}szl\'{o}~\surname{Botty\'{a}n}} \affiliation{Wigner RCP, RMKI, P.O. Box 49, H-1525, Budapest, Hungary}


\begin{abstract}
General expressions of the electron yield in \(^{57}\)Fe integral conversion
electron M\"{ossbauer} spectroscopy were derived depending on the glancing angle 
of the \(\gamma\) photons, on the source
polarization and on the isotopic abundance of the source and the absorber
(blackness effects) using an exponential escape function of the electrons
originating from all M\"{o}ssbauer-resonance-related processes. The present
approach provides a firm theoretical basis to determine the alignment and
direction of magnetization in the absorber. The intensity formulae were
justified by least squares fits of \(\alpha-^{57}\)Fe spectral intensities measured
in linearly and elliptically polarized source and absorber geometries. The
fits reproduce the experimentally set angles with high accuracy. Limits of
the current approach and its relation to other, less complete treatments in
the literature are discussed. 
\end{abstract}


\maketitle
\section{Introduction}
M\"{o}ssbauer spectroscopy has been very successful in determination of the
alignment and direction of magnetization of buried layers of Fe, the line
intensities being dependent on the angle between the propagation direction
of the \(\gamma\) photons from the source and the hyperfine field. Using linearly
polarized source we may determine alignment, and, using circularly polarized
radiation, also the sign of the magnetization component parallel to the
propagation direction \cite{Gonser66,Shtrikman69}. For this reason polarized
M\"{o}ssbauer sources were prepared \cite{Frauenfelder62, Szymansky96}. The
 so-called filter technique \cite{Szymansky96} provides almost single-line circularly
 polarized radiation, resulting in  relatively simple
spectra, but uses only a small part of the source intensity.
Application of a split multi-line source \cite{Shtrikman69,Gonser66,Frauenfelder62},
 however, results in more complex spectra, but
provides higher intensity.

Due to the relatively shallow escape depth of the conversion electrons, of typically
 5--10 keV initial energy, conversion
electron M\"{o}ssbauer spectroscopy (CEMS) is used for surface and thin film
investigations. Not even in the case of homogeneous
magnetization are conventional (perpendicular incidence) CEMS experiments appropriate to
determine an arbitrary direction of the magnetization.
The relative intensities do not depend on the alignment of the magnetization within
the sample plane, a frequent case with thin films, in which the shape anisotropy
constraints the magnetization in the sample plane. 

The theoretical description of transmission M\"{o}ssbauer spectroscopy is
available for quite general cases. However, in a CEMS\ experiment,  the
electron transport from the M\"{o}ssbauer nucleus to the sample surface is
rather complex which renders the calculation of the electron yield
difficult. There exist semi-empirical formulae (based on Monte-Carlo
simulations) providing the probability of the escape of an electron emitted
with a given energy at a given depth below the surface \cite{Liljequist78},
which can be built into the numerical algorithms, but seemingly do not allow
analytical solutions, which would provide more thorough insight. As a
result, most spectroscopists, when evaluating CEM spectra, neglect the use of such
escape functions and assume that the sample is very thin. In case of
perpendicular-incidence and non-enriched samples, this may be justified,
unlike for tilted incidence and for absorbers enriched in the M\"{o}%
ssbauer-active isotope, when polarization- and angle-dependent blackness
effects are appreciable. Even from otherwise quite general treatments of
multilayer structures, e.g. \Refonlinecite{NagyFero06} discussion of polarization and
blackness effects are missing. Papers that qualitatively elaborate such
effects, e.g. \Refonlinecite{Olszewski08}, leave room for improvement in regard
of generality and mathematical thoroughness. The opposite extreme is the grazing
incidence (approximately below \(3^\circ\)) CEM spectroscopy \cite{Irkaev93_2}, for which the theory is well
established, but, as we shall see, at such angles, a different 
approximation of the theory applies. Moreover, experimental feasibility of grazing incidence CEMS 
using radioactive sources is quite limited due to the fact, that the intensity
 is distributed in the complete solid angle, which
enables the use of less than a fraction of \(\approx10^{-5}\) of the source's activity.

In the last decade several instruments and methods were devised in our laboratory
for CEM polarimetry at arbitrary (except grazing) angle of incidence and
linear as well as elliptical polarization \cite{Tancziko04,Tancziko10-1,Tancziko09}. The analysis
of the measured spectra with these setups lead us to recognize, that
treatment of the blackness effect and polarization of the radiation in the
literature for conversion electron M\"{o}ssbauer spectroscopy at small glancing angles and of 
enriched samples in the resonant isotope is not satisfactory \cite{Tancziko10-1}. The present paper 
tries to fill this gap.
Here we derive the CEM spectral intensity in case of polarized (M\"{o}ssbauer) sources, 
when the \(\gamma\) photons fall on a homogeneously magnetized sample at arbitrary (except for 
grazing) angle of incidence. 
We will discuss in detail, under what conditions conversion electron absorption effects should
 be accounted for and, correspondingly, whether or not blackness effect will significantly
 influence the resonance line intensities and widths.
In particular, for \(^{57}\)Fe CEM spectroscopy, we will show that, due to the limited escape
depth of conversion electrons, CEM samples, even those prepared from enriched \(^{57}\)Fe,
can always be treated in the thin-sample limit, as long as the incidence of resonant photons is
 nearly perpendicular.
Conversely, for low- (but not grazing-) angle incidence an effective thickness can
be introduced so that CEM spectra correspond to transmission spectra measured on
resonant samples of this effective thickness if the resonance line polarization is properly 
taken into account in both cases.

\section{\label{sec:Theory} CEM spectra with the polarized \(\gamma\) photons incident on
 the absorber at an arbitrary angle}
In the following, concerning nuclear resonance we will follow the theoretical
considerations used in \Refonlinecite{DeakPRB96}.
 Here the case of a single, homogeneous layer will be considered.
Generalization for multilayers is straightforward (see Appendix~\ref{app:CemsMultiLayer}).

The number of detected electrons created by a photon beam of
energy \(E\) may be written as: 
\begin{equation}
\mathcal{N}_\text{CE}(D,E)\propto\int\limits_0^D\Tr\left(\tens{\rho}(z,E)
N(z)\tens{\sigma}^M(z,E)\right)W(z)\d z, \label{eq:Icems}
\end{equation} 
where \(D\) is the sample thickness, \(\tens{\rho}(z)\) is the density matrix
of the photons at a depth \(z\), \(N\) is the density of the scattering centres, 
\(\tens{\sigma}\) is the absorption cross section tensor and upper index \(M\)
 indicates the nuclear (M\"{o}ssbauer) contribution to the quantity. \(W(z)\)
is the probability of a conversion electron to be emitted following the absorption
of an incoming photon at depth \(z\) provided that it reaches the surface
of the sample. In the case of \textit{differential CEMS} (DCEMS) one should
write \(W(z,E_e)\), where \(E_e\) is the energy of the emitted electron, and
\(W(z,E_e)\) is the escape probability of an electron being emitted with
energy \(E_e\) \cite{Liljequist85}. For brevity, unless it may lead to
misunderstanding, we will omit the energy dependence in the formulae.

The transfer operator \(\tens{T}(z)\) defined by
\begin{equation}
\left|\psi(z) \right\rangle=\tens{T}(z)\left|\psi(0) \right\rangle \label{eq:TPsi}
\end{equation}
determines the wave at a depth \(z\) from the wave \(\left|\psi(z=0) \right\rangle\)
at the surface. Accordingly, for the density matrix we get
\begin{equation}
\tens{\rho}(z)=\tens{T}(z)\tens{\rho}(0)\tens{T}\herm(z), \label{eq:TRho}
\end{equation}
where \(\tens{T}\herm\) denotes the Hermitian adjoint operator of \(\tens{T}\).
After some permutation and substitution we may write
\begin{eqnarray}
&\mathcal{N}_\text{CE}(D,E)\propto\Tr \left(\tens{\rho}(0,E)\tens{J}(D,E)\right),
\quad\text{where} \label{eq:Ncems}\\
&\tens{J}(D,E)=\int\limits_0^D\tens{T}\herm(z,E)\tens{\sigma}^M(z,E)\tens{T}(z,E) 
N(z) W(z)\d z.\quad
\label{eq:Jcems}
\end{eqnarray}
For a single homogeneous layer \(\tens{T}(z)\) has the following form \cite{DeakPRB07}
 (see also Appendix~\ref{app:Transf}):
\begin{equation}
\begin{split}
\tens{T}(z)&=
\cosh\left(\imunit k_0\tens{m}z\sin\vartheta\right)(\tens{I}+\tens{R})\\
&+\sinh\left(\imunit k_0\tens{m}z\sin\vartheta\right)\tens{m}^{-1}(\tens{I}-\tens{R}),
\end{split} \label{eq:Tdefinition}
\end{equation}
where \(k_0\) is the wave number in vacuum, \(\vartheta\) is the glancing
angle of the incoming \(\gamma\)-beam, \(\tens{I}\) denotes the unity matrix, 
\(\left|\psi_\text{Reflected}(0) \right\rangle=\tens{R}\left|\psi(0) \right\rangle\) 
is the reflected wave (with \(\tens{R}\) being the reflectivity tensor) at the
surface of the sample. Here we introduced the tensor quantity \(\tens{m}\) with the definition
\begin{equation}
\tens{m}=\sqrt{\tens{I}+\frac{1}{\sin^2\vartheta}\tens{\chi}},\label{eq:mdef}
\end{equation}
where \(\tens{\chi}\) is the dielectric susceptibility tensor. For large glancing angles 
\(\vartheta,\) the reflected beam for \(\gamma\) photons is weak, i.e.\
\(\tens{R}\) is negligible compared to the unity matrix. Consequently 
\begin{equation}
\begin{split}
\tens{T}(z)&=\cosh\left(\imunit k_0\tens{m}z\sin\vartheta\right)+
\sinh\left(\imunit k_0\tens{m}z\sin\vartheta\right)\tens{m}^{-1}\\
&=\e^{\imunit k_0\tens{m}z\sin\vartheta}+
\sinh\left(\imunit k_0\tens{m}z\sin\vartheta\right)\left(\tens{m}^{-1}-\tens{I}\right).
\end{split}\label{eq:TdefinitionWithoutR}
\end{equation}
For multilayers the resulting transfer matrix is the product of single layer
 \(\tens{T}\) matrices of form (\ref{eq:TdefinitionWithoutR}).
For nearly perpendicular-incidence (\(\vartheta\approx\frac{\pi}{2}\)) the
matrix \(\tens{m}\) will be the index of refraction matrix
 \(\tens{n}=\sqrt{\tens{I}+\tens{\chi}}\approx \tens{I}+\frac{1}{2}\tens{\chi}\).
Since the dielectric susceptibility for \(\gamma\) photons is always small, the inverse of
the index of refraction matrix is also close to the unity matrix,
\(\tens{n}^{-1}\approx\tens{I}\), which leads to the approximation given
in \Refonlinecite{BlumeKistner}, namely:
\begin{equation}
\tens{T}(z)=\e^{\imunit k_0\tens{n}z}\approx\e^{\imunit k_0z}
\e^{\imunit \frac{1}{2}k_0\tens{\chi}z}.\label{eq:TSimplePerpendicular}
\end{equation}

For small glancing angles the approximation (\ref{eq:TSimplePerpendicular}) is
not valid. We may expand, however, \(\tens{m}\) and \(\tens{m}^{-1}\) into Taylor-series:
\begin{eqnarray}
\tens{m}&=&\tens{I}+\frac{1}{2\sin^2\vartheta}\tens{\chi}-
\frac{1}{8\sin^4\vartheta}\tens{\chi}^2+\dots\nonumber\\
\tens{m}^{-1}&=&\tens{I}-\frac{1}{2\sin^2\vartheta}
\tens{\chi}+\frac{3}{8\sin^4\vartheta}\tens{\chi}^2+\dots
\end{eqnarray}
In first order we get
\begin{equation}
\tens{T}(z)\approx\e^{\imunit k_0z\sin\vartheta}\e^{
\frac{\imunit}{2}k_0\tens{\chi}\textstyle\frac{z}{\sin\vartheta}}+\sinh\left(\imunit k_0\tens{m}z\sin\vartheta\right)\left(\tens{m}^{-1}-\tens{I}\right).\label{eq:Telsorend}
\end{equation}
Higher-order terms are negligible if the eigenvalues of \(\tens{\chi}\) fulfill
the condition \(|\chi_i|<\sin^2\vartheta\). For \(^{57}\text{Fe}\) on
resonance \((|\chi_i|\lessapprox0.01)\), therefore the first-order approximation
is valid down to about \(\vartheta\approx\)3--5\(^\circ\).

Unless for very small angles, \(\tens{m}\approx\tens{I}\) and therefore
\(\tens{m}^{-1}\approx\tens{I}\). Since in the following, we will not
consider grazing incidence and will limit our considerations to
\(\vartheta\gtrapprox 3^\circ\), we can neglect the second term
\(\sinh(\dots)(\tens{m}^{-1}-\tens{I})\) on the right side of  (\ref{eq:Telsorend}).

The absorption cross-section tensor for a single atom (molecule, etc.)
can be defined as \cite{Irkaev93_2,LandauAbszHatker} 
\begin{equation}
\tens{\sigma}= -\frac{\imunit k_0}{2N}\left(\tens{\chi}-\tens{\chi}\herm\right) \approx-\frac{\imunit k_0}{N}\left(\tens{n}-\tens{n}\herm\right).\label{eq:sigmaDef}
\end{equation}
In the following we assume, that the electronic contribution to the
susceptibility \(\chi^{e}\) is isotropic and independent of the energy
around the resonance, therefore
\begin{equation}
\tens{\chi}(E)=\tens{\chi}^M(E)+{\chi^e}\tens{I}.
\end{equation}

Substituting (\ref{eq:sigmaDef}) and the first term of (\ref{eq:Telsorend})
into  (\ref{eq:Jcems}), the CEMS intensity matrix \(\tens{J}\)  for a
homogeneous layer becomes:
\begin{widetext}
\begin{equation}
\begin{split}
\tens{J}(D)&=-\int_0^D \e^{-\imunit \frac{1}{2}k_0\left(
{\chi^e}^*-\chi^e\right)\textstyle\frac{z}{\sin\vartheta}}
 \e^{-\imunit \frac{1}{2}k_0{\tens{\chi}^M}\herm\textstyle\frac{z}{\sin\vartheta}}
\frac{\imunit k_0}{2}\left(\tens{\chi}^M-{\tens{\chi}^M}\herm\right)
 \e^{\imunit \frac{1}{2}k_0\tens{\chi}^M\textstyle\frac{z}{\sin\vartheta}}
W(z)\d z\\
&=-\sin\vartheta \int_0^D W(z)\e^{- N\sigma^e\textstyle\frac{z}{\sin\vartheta}} 
\frac{\partial}{\partial z}\left(\e^{-\imunit \frac{1}{2}k_0{\tens{\chi}^M}\herm
\textstyle\frac{z}{\sin\vartheta}}\e^{\imunit \frac{1}{2}k_0\tens{\chi}^M
\textstyle\frac{z}{\sin\vartheta}}\right)\d z,
\end{split}
\end{equation}
\end{widetext}
where \(*\) denotes complex conjugation.
Introducing the \(s=\imunit \frac{1}{2}k_0 z/\sin\vartheta\) notation and using
 the Baker\,--\,Campbell\,--\,Hausdorff formula \cite{Hairer06CBH} the product of
 the two matrix exponentials can be written as 
\begin{equation}
\begin{split}
\e^{-s{\tens{\chi}^M}\herm}\e^{s\tens{\chi}^M}=
\e&^{s\left(\tens{\chi}^M-{\tens{\chi}^M}\herm\right)-\frac{1}{2}s^2
\left[{\tens{\chi}^M}\herm,\tens{\chi}^M\right]}\\
&^{+\frac{1}{12}s^3 \left[\tens{\chi}^M-{\tens{\chi}^M}\herm,
\left[{\tens{\chi}^M}\herm,\tens{\chi}^M\right]\right]+\cdots}.
\label{eq:BakCampHaus}
\end{split}
\end{equation}
For the M1 nuclear transitions, like in \(^{57}\text{Fe}\), for almost all
 practical cases the \(\left[\tens{\chi}^M,{\tens{\chi}^M}\herm\right]=0\)
 condition is fulfilled  with a good approximation \cite{Spiering84}. Otherwise
 one has to estimate the contribution of higher-order commutator terms in
 (\ref{eq:BakCampHaus}). In the following, we assume that \(\left[\tens{\chi}^M,
{\tens{\chi}^M}\herm\right]=0\) is fulfilled, therefore we have
\begin{eqnarray}
&\tens{J}&(D,E)=\nonumber\\
&=&-\sin\vartheta \int_0^D W(z)\e^{- N\sigma^e\textstyle\frac{z}{\sin\vartheta}}
\frac{\partial}{\partial z}\e^{- N\tens{\sigma}^M(E)
\textstyle\frac{z}{\sin\vartheta}}\d z\quad\quad\label{eq:JcemsSimple}\\
&=&\int_0^D W(z) N\tens{\sigma}^M(E)\e^{- N\left(\sigma^e\tens{I}+
\tens{\sigma}^M(E)\right)\textstyle\frac{z}{\sin\vartheta}}\d z.  \label{eq:JcemsSimpleX}
\end{eqnarray}
Not regarding the number of approximations we had to make in order to obtain
eq.~(\ref{eq:JcemsSimple}), the result may seem to be commonsensical: the
\(\sin \vartheta\) term in front of the integral simply takes the reduced cross section
of a tilted sample into account. The number of resonantly absorbed photons at
depth \(z\) is given by the derivative term inside the integral. For small angles, the
photons travel a longer path \(z/\sin\vartheta\) to reach depth \(z\) and therefore
they get absorbed more likely near the surface as compared to the case of 
perpendicular-incidence. However, one should keep in mind, that eq.~(\ref{eq:JcemsSimple}) is only
valid for not too small angles, when the linear approximation in eq.~(\ref{eq:Telsorend})
is valid and the reflection is negligible and \(\left[\tens{\chi}^M,{\tens{\chi}^M}\herm\right]=0\).

Until now we did not take an essential feature of the M\"{o}ssbauer effect into account,
namely the moving source. The density matrix of the \(\gamma\)-beam emitted by
the M\"{o}ssbauer source that moves with velocity \(v\) can be written as
\begin{equation}
\begin{split}
\tens{\rho}(z=0,E,v)&=\sum\limits_i\tens{\rho}_i^{\text{pol}}\mathcal{L}_i(E,v)\\
&=\sum\limits_i\tens{\rho}_i^{\text{pol}}\frac{\Gamma}{2\pi}
\frac{1}{\left(E-E_i+\frac{v}{c}E_i\right)^2+\frac{\Gamma^2}{4}},
\end{split}
\end{equation}
where \(\tens{\rho}_i^\text{pol}\) characterizes the intensity ratio and polarization
 of the \(i\)-th line at position \(E_i\) in the energy spectrum, \(\mathcal{L}_i(E,v)\)
 is the (Lorentzian) shape function of the \(i\)-th line, \(\Gamma\) is the line width and
 \(c\) is the speed of light. Disregarding the background the measured CEMS intensity, as a
 function of the energy, i.e.\ of  the M\"{o}ssbauer drive velocity can be written as:

\begin{equation}
I_\text{CEMS}(D,v)= f_S\mathcal{I}\sum\limits_i\Tr \left(\tens{\rho}_i^{\text{pol}}
\int \mathcal{L}_i(E,v) \tens{J}(D,E)\d E\right),
\end{equation}
where \(f_S\) is M\"{ossbauer}\,--\,Lamb factor of the source and \(\mathcal{I}\) 
is the total intensity of the source incident on the sample.

In order to calculate the CEM spectra the knowledge of the electron escape function
 \(W(z)\) is required. In the following, we consider \(^{57}\)Fe CEM spectroscopy.
 The same treatment may be applied to CEMS on other isotopes. Liljequist derived
 semi-empirical formulae based on his Monte-Carlo simulations \cite{Liljequist78,Liljequist98}.
 Instead of depth, he used the `equivalent depth in iron' defined as 
\begin{equation}
t=\frac{\varrho}{\varrho_\text{Fe}} z,\label{eq:tDefinition}
\end{equation}
where \(\varrho\) is the density of the sample and \(\varrho_\text{Fe}\) the density
 of natural iron at ambient conditions.
For a homogeneous layer, the escape function can be well approximated with an exponential, 
\begin{equation}
W(z)=W_0\e^{\textstyle-\frac{t(z)}{\tau}}=W_0\e^{\textstyle-z
\frac{\varrho}{\varrho_\text{Fe}\tau}}=W_0\e^{\textstyle-\frac{z}{\zeta(\varrho)}},
\end{equation}
where  \(W_0\approx1.155\)  and {\(\tau\approx\) 540~Fe\AA}. The more
 general escape functions are shown in Appendix~\ref{app:Lilj}. The exponential
 approximation is advantageous in the parameter fitting routines, since the computer code
 based on the exponential escape function is by two orders of magnitude faster
 than the one based on exact formulae, while providing a rather good approximation.
 Moreover, the exponential approximation is easier to handle analytically in some
 special cases. Indeed, with (\ref{eq:JcemsSimpleX}) and neglecting
 \(\sigma^e\)\footnote{This usually can be done near the resonance and
 for the glancing angles examined here. Even if it is not the case, this exponential
 term could be taken into account formally just by replacing \(\frac{1}{\zeta}\) 
with \(\frac{1}{\zeta}+\frac{N\sigma^e}{\sin\vartheta}\).}, we have
\begin{equation}
\begin{split}
&I_\text{CEMS}(D,v)= a
\sum\limits_i\Tr 
\left[\tens{\rho}_i^{\text{pol}}
\int \mathcal{L}_i(E,v) \zeta N\tens{\sigma}^M(E)\sin\vartheta \right.\\
&\times\left.\left(\zeta N\tens{\sigma}^M(E)+\tens{I}\sin\vartheta\right)^{-1}
\left(\tens{I}-\e^{- D\left(\frac{N\tens{\sigma}^M(E)}{\sin\vartheta}
+\frac{1}{\zeta}\tens{I}\right)}\right) \d E\right],
\end{split}\label{eq:ICemsVeg}
\end{equation}
where \(a=W_0 f_S \mathcal{I}\).

First let us consider the unpolarized case. Applying the formula of effective thickness 
as \(t=D\varrho/\varrho_\text{Fe}=DN/N_\text{Fe},\) and similarly
 \(\tau=\zeta N/N_\text{Fe}\) and the absorption coefficient (for natural iron at
 ambient conditions) as \(\alpha^M=N_\text{Fe}\sigma^M\), the exponent in
 eq.~(\ref{eq:ICemsVeg}) can be transformed as follows:
\begin{equation}
-D\left(\frac{N\sigma^M(E)}{\sin\vartheta}+\frac{1}{\zeta}\right)=
-\alpha^M t\frac{\tau\alpha^M+\sin\vartheta}{\tau\alpha^M\sin\vartheta}.
\end{equation}
It may be convenient to interpret the quantity
\begin{equation}
t^\prime(E,\vartheta)=t\frac{\tau\alpha^M(E)+\sin\vartheta}{\tau\alpha^M(E)\sin\vartheta}
\end{equation}
as a corrected effective thickness \cite{Tancziko10-1}. 

In the polarised case for small angles (\(\sin\vartheta\approx 0\)) we have
 \(\zeta N\tens{\sigma}^M(E)\left(\zeta N\tens{\sigma}^M(E)+
\tens{I}\sin\vartheta\right)^{-1}\approx\tens{I}\) and therefore 
\begin{equation}
\begin{split}
I_\text{CEMS}(D,v)= a \sin\vartheta
&\left[\tens{I}-\sum\limits_i\Tr 
\left(\tens{\rho}_i^{\text{pol}}
\int \mathcal{L}_i(E,v)\right.\right.\\&\left.\left. \times
\e^{- N\tens{\sigma}^M(E)\frac{D}{\sin\vartheta}} 
\d E \right)\right],
\end{split}
\end{equation}
which formally corresponds to the thick sample case in absorption M\"{ossbauer}
 spectroscopy, as it was conjectured in \Refonlinecite{Olszewski08}. We should
 not forget, however, that  beside the above mentioned constraints in deriving
 (\ref{eq:JcemsSimple}) and (\ref{eq:ICemsVeg}), this approximation is not valid
 for grazing angles, when the reflections due to the electronic contribution to the
 susceptibility are not negligible.

At nearly perpendicular-incidence, when \(\sin\vartheta\approx 1\) we expand the
 exponent in Taylor series up to the second order:
\begin{equation}
\begin{split}
&\left(\tens{I}-\e^{- D\left(N\tens{\sigma}^M(E)+
\frac{1}{\zeta}\tens{I}\right)}\right)\approx\\ 
&D\left(N\tens{\sigma}^M(E)+\frac{1}{\zeta}\tens{I}\right)+
\frac{1}{2}D^2\left(N\sigma^M_i(E)+\frac{1}{\zeta}\right)^2+... \label{eq:Condition2}
\end{split}
\end{equation}
In case the second term is negligible compared to the first one, i.e.
\begin{equation}
\frac{1}{2}D^2\left(N\sigma^M_i(E)+\frac{1}{\zeta}\right)^2
 \ll D\left(N\sigma^M_i(E)+\frac{1}{\zeta}\right),
\end{equation}
where \(\sigma_i\) is the \(i\)-th eigenvalue of \(\tens{\sigma}\), and therefore
 the intensity expression is

\begin{equation}
I_\text{CEMS}(D,v)= a D
\sum\limits_i\Tr 
\left(\tens{\rho}_i^{\text{pol}}
\int N\tens{\sigma}^M(E)\mathcal{L}_i(E,v)
\d E\right),\label{eq:ICemsThin}
\end{equation}
which formally corresponds to the thin sample case in M\"{ossbauer} spectroscopy,
 as discussed in \Refonlinecite{Olszewski08}.

From the above, the condition of the thin absorber limit in perpendicular-incidence is
\begin{equation}
D \ll \frac{2}{N\sigma^M_\text{max}+\frac{1}{\zeta}}, \label{eq:ThinCondition}
\end{equation}
where \(\sigma^M_\text{max}\) is the maximum eigenvalue of the absorption cross
 section tensor. This condition means, that \(D\) is the characteristic
information depth of the surface \ of the sample for CEM\ spectroscopy; no
significant number of electrons will reach the sample surface from deeper
layers of the sample and/or the photons will not reach a greater perpendicular depth in the sample.

For multilayers we may follow the same steps (see Appendix~\ref{app:CemsMultiLayer}),
 the main difference and question will be the electron escape function, which may
 become quite complex. The conversion electrons emitted from atoms in different
 layers should be summed up, taking into account the proper escape function(s),
 which probably may be determined as some examples of \Refonlinecite{NagyFero06}
 show, but this is out of the scope of the present paper.

\section{Thick source case}
Blackness effects in M\"{o}ssbauer spectroscopy are not limited to the absorber. 
Due to the possible self-absorption of the emitted photons, additional broadening
 may arise \cite{Margulis61} or, in case of a split multi-line source, the relative
 intensities of the emitted lines of the M\"{o}ssbauer source may considerably
 be modified, provided that \(\gamma\)-emission occurs relatively deep below
 the surface of the source and the substrate of the source contains large amount
 of the resonant isotope. In the commercial M\"{o}ssbauer sources the precursor
 of the resonant isotope is diffused into the surface layer
 of the substrate. When using such sources in perpendicular emission geometry,
 blackness effect rarely occur. However, the relative intensities of a split multiline 
source of \(\alpha\)-Fe substrate at tilted exit of the \(\gamma\) photons, which
 is studied experimentally and discussed below are considerably modified by the
 blackness effects. In the following, we discuss the case of the thick source. 

Following the same logic as in deriving (\ref{eq:TRho}), the density matrix of 
the photons emitted by the M\"{o}ssbauer source may be written as
\begin{equation}
 \tens{\rho}(E)\propto\int\limits_0^{D_s}\tens{\mathfrak{T}}(z,D_s-z,E)
\tens{\sigma}^M(E)\tens{\mathfrak{T}}\herm(z,D_s-z,E) \mathcal{N}_s(z)\d z,\label{eq:ThickSource}
\end{equation}
where \(D_s\) is the physical thickness of the substrate of the source and 
\(\mathcal{N}_s(z)\) is the number of the nuclei emitting the resonant
 photons at depth \(z\), \(\tens{\mathfrak{T}}(z,D_s-z)\) is the transfer operator
 for the source from depth \(z\) through a layer of thickness \(D_s-z\) to the surface of
 the M\"{o}ssbauer source. Generally, the source may not be regarded as a
 homogeneous layer in this case. Therefore \(\tens{\mathfrak{T}}\) should be calculated
 somewhat differently, than \(\tens{T}\) in (\ref{eq:TdefinitionWithoutR}).
 In principle it may be derived as a product of \(\tens{T}\) matrices of consecutive infinitesimally thin layers.
 In the present case, it may be assumed, that \(\tens{\chi}^M\) only changes due to
  the density, the number \(\mathcal{N}_{^{57}\text{Fe}}\) of resonantly
 absorbing iron changes, i.e.\ \(\tens{\chi}^M_s(z)=\tens{\chi}^M_0
 \mathcal{N}_{^{57}\text{Fe}}(z)/(\int_0^{D_s}\mathcal{N}_{^{57}\text{Fe}}(z)\d z), \) 
where \(\tens{\chi}^M_0\) is the average susceptibility in the sample. In that case, 
instead of such a product we may express \(\tens{\mathfrak{T}}\) using the
 only the first exponential terms in formulae (\ref{eq:Telsorend}), i.e.\
\(\tens{T}(z)\approx\e^{\imunit k_0z\sin\vartheta}\e^{\imunit
\frac{1}{2}k_0\tens{\chi}\textstyle\frac{z}{\sin\vartheta}}\) available for
 homogeneous layers, from which we get:
\begin{equation}
\tens{\mathfrak{T}}(z,D_s-z)=\e^{\imunit k_0\left(D_s-z\right)\sin\vartheta_s}\e^{\imunit
\frac{1}{2}k_0\chi^e_s\textstyle\frac{D_s-z}{\sin\vartheta_s}}
\e^{\imunit
\frac{1}{2}k_0\tens{\chi}^M_0\textstyle\frac{\mathcal{Z}}{\sin\vartheta_s}},
\end{equation}
where \(\vartheta_s\) is the tilt angle of the source, \(\chi^e_s\) is the electronic contribution
to the susceptibility for the source (assumed to be constant) and
\begin{equation}
\mathcal{Z}=\frac{\int\limits_z^{D_s}\mathcal{N}_{^{57}\text{Fe}}(z)\d z}
{\int\limits_0^{D_s}\mathcal{N}_{^{57}\text{Fe}}(z)\d z}D_s.
\end{equation}
Due to the preparation procedure, it is correct to assume that \(\mathcal{N}_s(z)\) follows
 a diffusion profile, i.e.\ an expression containing an error function.

\section{Experiments and theory}

CEM polarimetric experiments were performed using \(\alpha\)-iron foils, a natural
Fe foil of 15~\(\mu\)m thickness as the substrate of the \(^{57}\text{Co}(\alpha \text{-Fe})\)
source and a \(^{57}\text{Fe}\) foil of 20~\(\mu\)m thickness as the sample. A
detailed description of these experiments was published in \Refonlinecite{Tancziko10-1}.
 The experimental spectra are the same here as in \Refonlinecite{Tancziko10-1}, but here we use a different
model function. The emphasis here is being on the verification of the theory
derived in the present paper.

\begin{figure}[ht]
\includegraphics[clip,width=75mm]{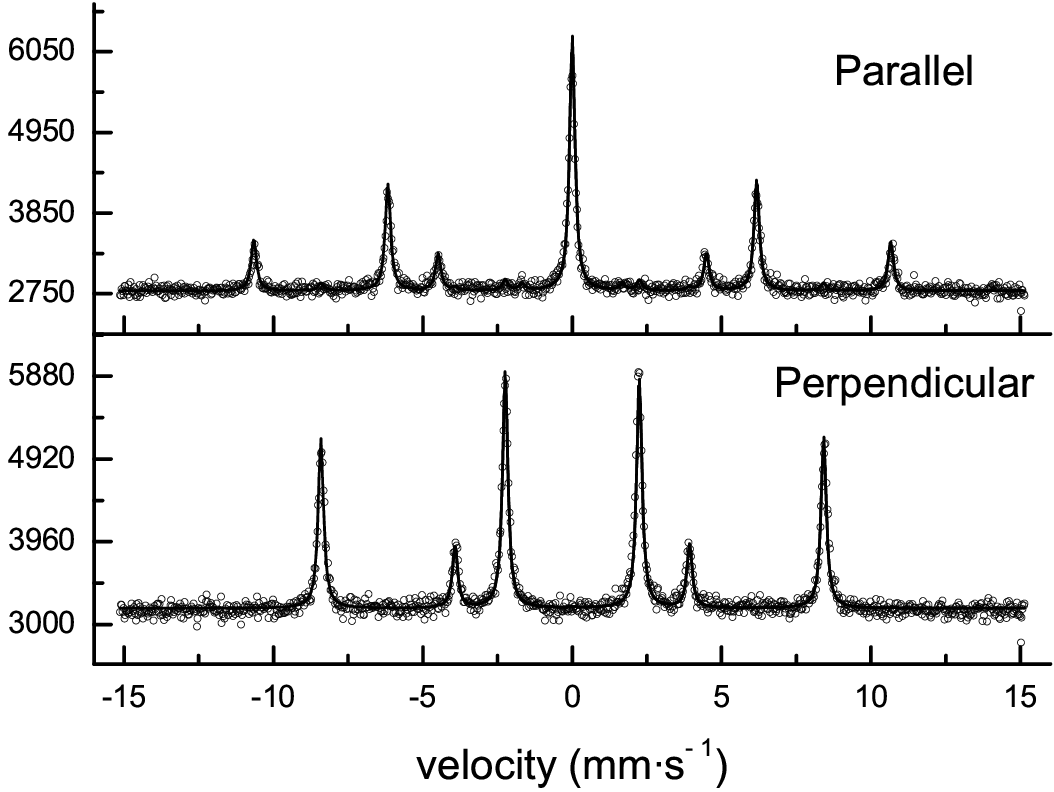}
\caption{\label{fig:LinOrthogonal} Results of linear polarimetric measurements (circles) and their fits (full lines) with
 parallel and perpendicular source and sample magnetizations. The \(\gamma\)-beam is perpendicular
 to the plane of the source and of the sample. The magnetizations are constrained in the plane
 of the source and sample, respectively.}
\end{figure}

Linear polarimetric experiments were performed in perpendicular-incidence geometry,
in which the external magnetic fields on the source (270~mT) and on the 
sample (400~mT) were magnetized perpendicular to the propagation direction.
The results of two such measurements are shown in Fig.~\ref{fig:LinOrthogonal}.\
where the magnetic fields on the source and sample are parallel and perpendicular to
 each other, respectively.

\begin{figure}[ht]
\includegraphics[clip,width=75mm]{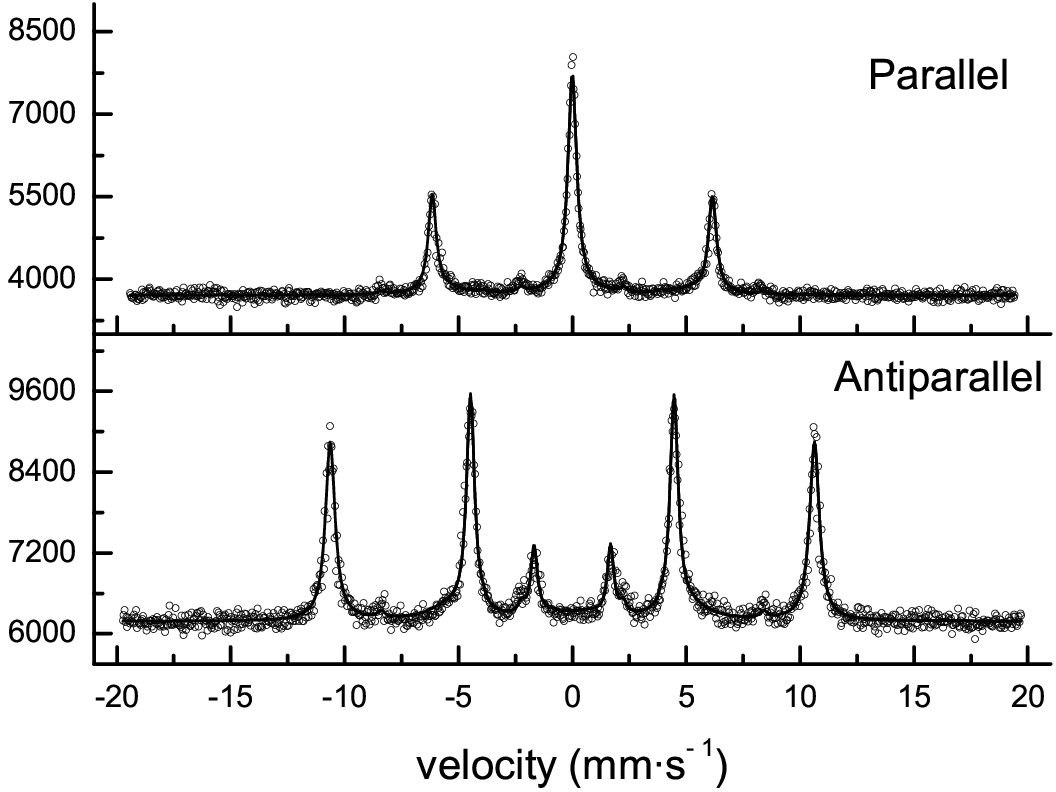}
\caption{\label{fig:CircAntiparallel} Results of elliptical polarimetric measurements (circles) and their fits (full lines)
 with parallel and antiparallel source and sample magnetizations. The \(\gamma\)-beam
 enclosed \(\vartheta_s=10^{\circ}\) with the plane of the source and \(\vartheta=5^{\circ}\)
 with the sample. The external magnetic fields were parallel to the lines of intersection of the 
scattering plane and the planes of the sample and the source, respectively.}
\end{figure}

Elliptical (close to circular) polarimetric experiments were performed with
a tilted source and a tilted sample, both magnetized  close to their planes the
magnetization vectors and the wave vector laying in the same plane. Their
planes enclosed \(\vartheta_s=10^{\circ}\) (source) and \(\vartheta=5^{\circ}\) (sample)
 with the propagation direction. The results of two such CEM experiments are
 shown in Fig.~\ref{fig:CircAntiparallel}.\ where the external magnetic fields
 on the source (230~mT) and on the sample (100~mT) were parallel and antiparallel to each other, respectively.

Having implemented the theory outlined in the former sections, measured
spectra were fitted using the general fitting program \textit{FitSuite} \cite{FitSuiteRef}.
The  positions and intensities of the M\"{o}ssbauer lines (i.e.\ the contribution
 of the resonant absorption to \(\tens{\sigma}(E)\)) were calculated solving the
 eigenvalue problem of the corresponding hyperfine Hamiltonian. The theory described
in \Refonlinecite{Spiering00} and implemented in the program EFFI \cite{EffiRef}
 were adopted in \textit{FitSuite} and
modified to include the theoretical considerations presented above, namely
eq.~(\ref{eq:ICemsVeg}).

\begin{table}[ht]
\caption{Essential fitted parameters, linear polarimetry. \(\vartheta_\text{B,source}\)
 and \(\vartheta_\text{B,sample}\) are the angles enclosed by the hyperfine field and
 the \(z\) axis (propagation direction of the \(\gamma\) photons). \(\varphi_\text{B,source}\) and
 \(\varphi_\text{B,sample}\) are the azimuthal angles measured around
 the \(z\) axis from the \(x\) axis being in the plane of incidence. \label{tabl:LinFitParTable}}
\begin{tabular}{lcr}
\hline
Parameter & Fitted & Expected\\
\hline
\(|\mathbf{B}|\) & 33.05 T \(\pm\) 0.02 T& --\\
\(\vartheta_\text{B,source}= \vartheta_\text{B,sample}\) & 90\(^\circ\) (fixed)\\
\(\varphi_\text{B,source}\) & 0\(^\circ\) (fixed)\\
\hline
Parallel case:\\
\(\varphi_\text{B,sample}-\varphi_\text{B,source}\) & \textcolor{blue}{-0.2\(^\circ\) \(\pm\) 0.8\(^\circ\)}& 0\(^\circ\)\\
\hline
Perpendicular case:\\
\(\varphi_\text{B,sample}-\varphi_\text{B,source}\) & \textcolor{blue}{89.8\(^\circ\) \(\pm\) 0.8\(^\circ\)}& 90\(^\circ\)\\
\hline
\end{tabular}
\end{table}
In the linear polarimetric case the \(\gamma\)-beam is perpendicular  to the source and
 sample planes, therefore no thickness and angle dependence is expected according to
 (\ref{eq:ICemsThin}). The magnitude and orientations of the hyperfine fields (see
 Table~\ref{tabl:LinFitParTable}.) were found to agree quite well with the experimental
 values, set by the direction of the external magnetic fields. In this case, the
\(\vartheta\) tilt angles of the sample and source were not fitted.
These spectra depend on the difference \(\Delta\varphi\)
 \((=\varphi_\text{B,sample}-\varphi_\text{B,source})\) of
 the azimuthal angles of the hyperfine fields in the sample and the source, therefore \(\Delta\varphi\)
 was the fit parameter. The two spectra were simultaneously fitted using the constraint
\(\Delta\varphi^\text{perpendicular}=\Delta\varphi^\text{parallel}+90^\circ.\) Indeed, the
 perpendicular setup was obtained from the parallel one by the rotation of the 
source polarizer magnet by 90\(^\circ.\)
To reproduce the expected angles we had to assume that the source was not completely polarized, 
and that the emitted beam had a degree of polarization \(P = 0.98 \pm 0.007.\)
As the source was not saturated (As the coercive field of bulk bcc Fe is known not to be able fully saturate in plane an alpha-Fe foil and to saturate the last few per cents magnetic fields in the order of 1~T are needed. \footnote{R.~R{\"{o}}hlsberger, private communication; our unpublished results}), this is expectable, and the value of \(P\) agrees well with the value 
\(P = 0.96\) obtained in a different experiment in \Refonlinecite{Tancziko04}. The partially polarized source was 
modelled by using an additional site in the source with orthogonal \((\varphi_\text{B,source} =90^\circ)\)
hyperfine field of \((1-P)/(1+P)\) times smaller intensity than that of the `main component'. This results in a source spectrum of the desired degree of polarization and, as far as coherent scattering can be neglected, which is certainly the case as long as blackness effect plays no significant role, the contributions of different sites of the source can be added up incoherently.
Indeed, according to our simulations the change of the sample thickness gives 
rise to the change of intensity, but the lineshape of the spectra remains unchanged.

\begin{figure}[ht]
\includegraphics[clip,width=75mm]{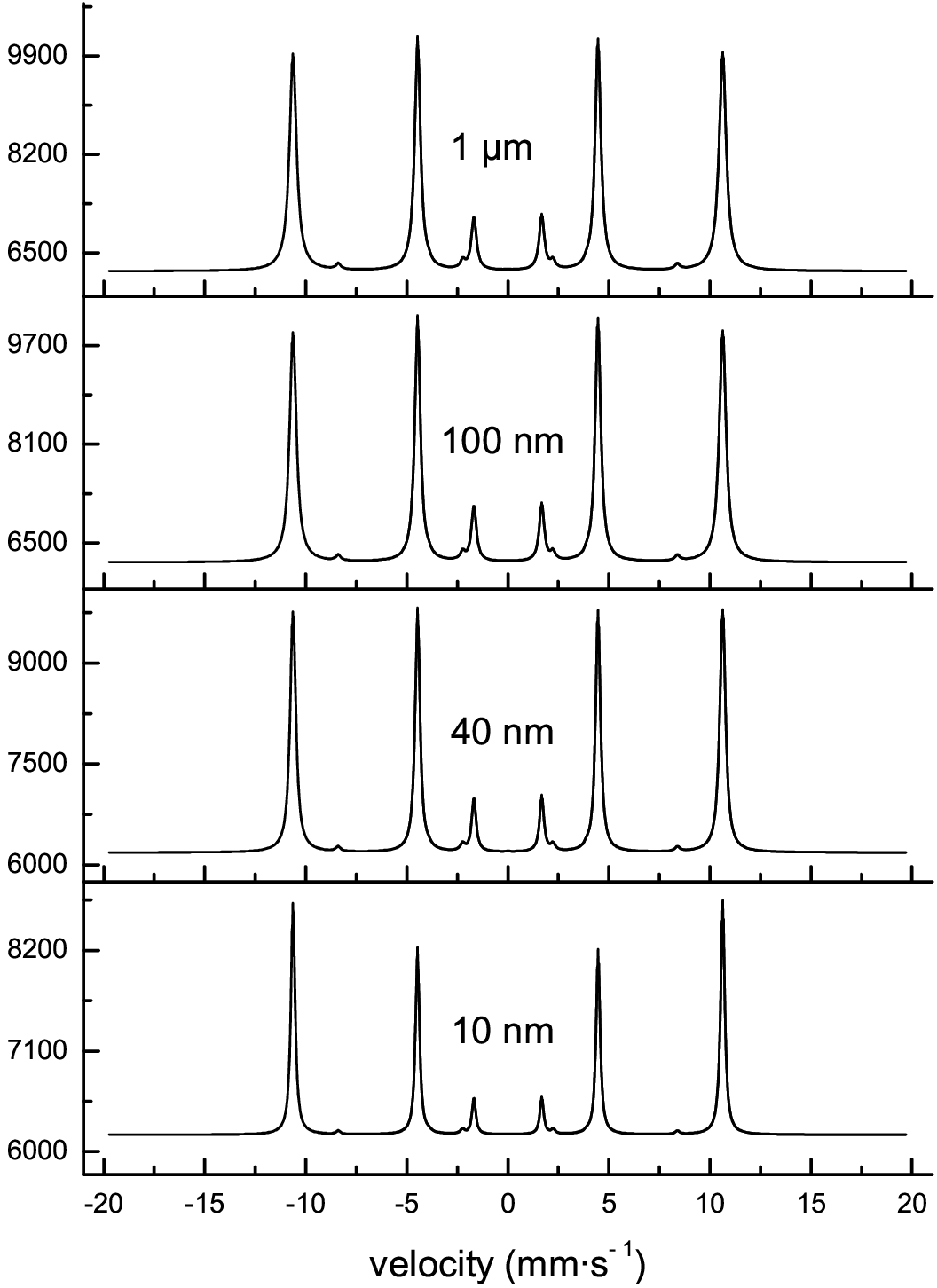}
\caption{\label{fig:VastFuggAnti} Demonstration of the thickness dependence of the blackness effect. Results
 of elliptical polarimetric simulations with antiparallel source and sample magnetizations
 for different sample thicknesses \((D= 10\)~\(\mathrm{nm},\) \dots, \(1\)~\(\mathrm{{\mu}m})\). Other parameters 
(see Table~\ref{tabl:CircFitParTable}) are the same as in the
 corresponding fitted spectrum plotted in Fig.~\ref{fig:CircAntiparallel}, but the additional site due
 to the superparamagnetic iron-oxide on sample surface was not taken into account and it was asssumed to have a thin source.}
\end{figure}

\begin{figure}[ht]
\includegraphics[clip,width=75mm]{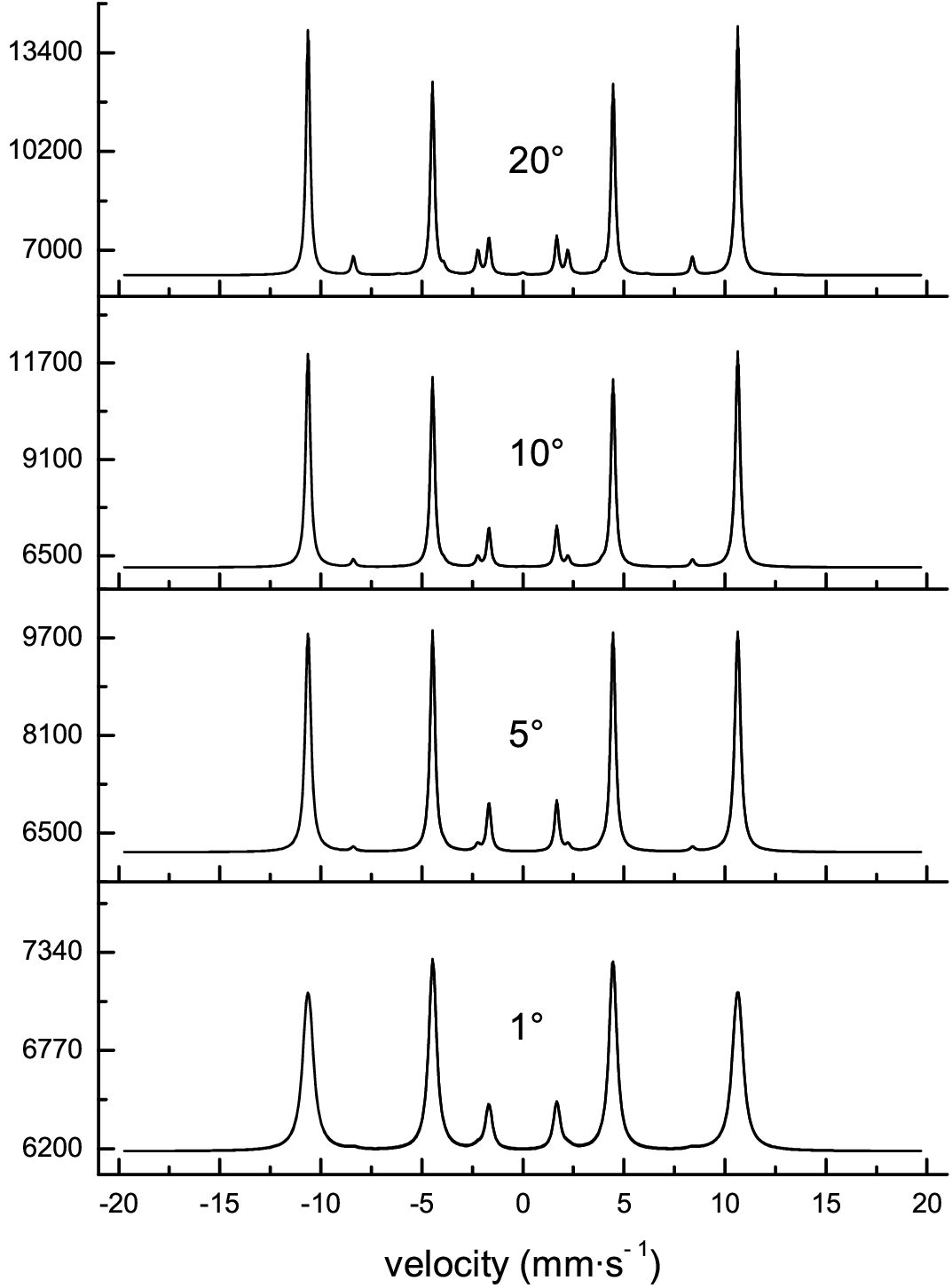}
\caption{\label{fig:SzogFuggAnti} Demonstration of the angular dependence of the blackness effect. Results of
 elliptical polarimetric simulations with antiparallel source and sample magnetizations
 for \(D\)=40~nm thick samples tilted in different angles \((\vartheta = 1^\circ,\) \dots, \(20^\circ)\). Other
 parameters (see Table~\ref{tabl:CircFitParTable}) are the same as in the
 corresponding fitted spectrum plotted in
 Fig.~\ref{fig:CircAntiparallel}, but the additional site due
 to the superparamagnetic iron-oxide on sample surface was not taken into account and it was asssumed to have a thin source.}
\end{figure}

\begin{table}[ht]
\caption{Essential fitted parameters, elliptical polarimetry. \(\vartheta_\text{B,source}\)
 and \(\vartheta_\text{B,sample}\) are the angles enclosed by the hyperfine field and
 the \(z\) axis (propagation direction of the \(\gamma\) photons). \(\varphi_\text{B,source}\) and
 \(\varphi_\text{B,sample}\) are the azimuthal angles measured around
 the \(z\) axis from the \(x\) axis being in the plane of incidence. The magnetizations are
assumed to lie in the source and sample planes, therefore (during fitting) \(\vartheta_\text{B,sample}\) and 
\(\vartheta_\text{B,source}\) were assumed to be identical with the corresponding tilt angles \(\vartheta\) and \(\vartheta_s\), respectively. \label{tabl:CircFitParTable}}
\begin{tabular}{lcc}
\hline
Parameter & Fitted & Expected\\
\hline
Source tilt angle \((\vartheta_s)\) & 10.1\(^\circ\) \(\pm\) 0.7\(^\circ\)& 10\(^\circ\)\\
Sample tilt angle \((\vartheta)\) & 4.97\(^\circ\) \(\pm\) 0.04\(^\circ\)& 5\(^\circ\)\\
\hline
\(|\mathbf{B}_\text{source}|\) & 32.866 T \(\pm\) 0.017 T&\,--\,\\
\(\vartheta_\text{B,source}\) & = \(\vartheta_s\) & 10\(^\circ\)\\
\(\varphi_\text{B,source}\) &  0\(^\circ\) (fixed)\\
\hline
\(|\mathbf{B}_\text{sample}|\) & 33.082 T \(\pm\) 0.002 T&\,--\,\\
Parallel case:\\
\(\vartheta_\text{B,sample}\) & = \(\vartheta\) & 5\(^\circ\)\\
\(\varphi_\text{B,sample}\) & 0\(^\circ\) (fixed) \\
\hline
Antiparallel case:\\
\(\vartheta_\text{B,sample}\) & = \(\vartheta\) & 5\(^\circ\)\\
\(\varphi_\text{B,sample}\) & 180\(^\circ\) (fixed)\\
\hline
\end{tabular}
\end{table}
This is not the case for the nearly circular polarimetric experiments, where the blackness
 effect plays an important role and therefore the intensity ratios strongly depend on
 the thickness and tilt angle, as it can be seen in Figs.~\ref{fig:VastFuggAnti}-\ref{fig:SzogFuggAnti}.
Without this effect, the intensity ratios of the spectra of Fig.~\ref{fig:CircAntiparallel}.\
 could not be explained either.
The measured spectra and hyperfine field orientations (Table~\ref{tabl:CircFitParTable}) could be
reproduced relatively well if the degree of polarization of the source was \(P= 0.924 \pm 0.010\).
This is worse, than in the linear polarization case, but it may be explained with
the incomplete saturation by the external magnetic field of 230~mT and by the experimental geometry: tilted
sample and source. The partial polarization of the source here was taken into account also by an additional site,
but here its hyperfine field was chosen antiparallel (as that belongs to the orthogonal polarization). In this case, because of the thick tilted source, this may not result completely
correctly the desired polarization degree, but it can be used, as in an experiment, \(P\) is just an additional calibration parameter. The
experimentalist would like to determine the orientation of the magnetization in the sample.

In fitting the circular polarimetric spectra at a glancing angle of
 \(\vartheta =5^{\circ}\), when the perpendicular information depth \(D\) is
much smaller, than in the perpendicular-incidence case. 
We had to take into account an additional singlet site with isomer shift 
(0.136 \(\pm\) 0.070) \(\mathrm{mm}\cdot\mathrm{s}^{-1}\),
 effective thickness (31.5 \(\pm\) 1.9) \(\mathrm{Fe}^{57}\text{\AA}\) with
 Lorentzian broadening of (17.5 \(\pm\) 1.4)\(\times \Gamma^\text{nat}\)
 (\(\Gamma^\text{nat}\)=0.097 \(\mathrm{mm}\cdot\mathrm{s}^{-1}\) is
 the width of the natural line broadening) which  is most probably due to a small
 amount of superparamagnetic iron oxide on the surface of the absorber.
In case of linear polarimetric measurements this component did not appear, since
 in perpendicular-incidence its contribution to the measured spectrum was much smaller.

The fits of elliptical polarimetric experiments, shown in \Refonlinecite{Tancziko10-1}
provided two times larger values for \((\vartheta_B)\) angles, which determine the tilt
of the magnetization, than the values available here in Table~\ref{tabl:CircFitParTable}
and what is expected from the experimental arrangement.
This clearly shows, that, instead of the simple model used in \Refonlinecite{Tancziko10-1}, 
the angular dependence, blackness and polarization effects are to be taken into account in 
a correct evaluation of CEM spectra, as outlined above.

\section{Summary}
In this paper it was examined, how the CEM spectra may be calculated for thick tilted single or
 multilayer samples measured by CEM polarimetry in order to determine the magnitude
 and direction of layer magnetization with high accuracy. We showed, what conditions are to be
fulfilled to justify the different approaches in evaluation: 
CEM spectra
may be regarded, as transmission M\"{o}ssbauer spectra (except for the change of absorption to emission lines): \begin{itemize}
 \item of a thin sample, if the \(\gamma\) photons from the source are incident on the sample
 at a large glancing angle or perpendicularly and the condition (\ref{eq:ThinCondition}) is fulfilled;
 \item of a thick sample, if the \(\gamma\) photons from the source are incident on the sample
 at a small glancing angle, which is still greater than what we usually call grazing incidence.
\end{itemize}
These statements are verified for electron escape of a single exponential form.
 Furthermore, we may not regard the CEM spectra either as a thin or a thick M\"{o}ssbauer
 spectrum for angles between the two limiting cases, and in case of grazing incidence. We may
 use for calculations the equations: (\ref{eq:Ncems}) and (\ref{eq:JcemsSimple}) for intermediate
angles (\(3^\circ\lessapprox\vartheta\leqslant90^\circ\)); (\ref{eq:Ncems}), (\ref{eq:Jcems}) and (\ref{eq:Tdefinition}) in
 case of grazing incidence and small glancing angles (\(\vartheta\lessapprox3^\circ\)).

We made numerous simulations and data fitting, using the FitSuite code, and we found
good agreement between theoretical and experimental results.
\appendix

\section{\label{app:CemsMultiLayer} CEM spectra in case of multilayers} 
Let us assume a sample of \(n\) homogeneous layers, and use the following notations:
\begin{itemize}
\item \(D_i\) is the depth \(z\) at which the interface of the \(i\)-th and the \(i+1\)-th
 layer can be found, \(D_0=0\) and \(D_n=D\),
\item \(\tens{\mathcal{T}}_i(z)\) is the transfer operator in the range
 \((D_{i-1}\leqslant z \leqslant D_i)\) of the two bounding interfaces of the \(i\)-th layer, 
\(\tens{\mathcal{T}}_i(D_{i-1})=\tens{I}\),
\item \(\tens{\sigma}_i^M\) and \(\sigma_i^e\) are the part of the absorption cross
section tensor belonging to M\"{o}ssbauer effect and the electronic contribution to the 
absorption cross section for the \(i\)-th layer, respectively,
\item \(N_i\) is the density of the M\"{o}ssbauer resonant nuclei in the \(i\)-th layer.
\end{itemize}
Using this notation the number of detected electrons may be written as:
\begin{equation}
\begin{split}
&\mathcal{N}_\text{CE}(D,E)\propto\sum\limits_{i=1}^n
\Tr \left(\tens{\rho}_i(E)\tens{J}_i(E)\right),\quad\text{where}\\
&\tens{\rho}_i(E)=\tens{T}(D_{i-1},E)\tens{\rho}(0,E)\tens{T}\herm(D_{i-1},E)\\
&\tens{J}_i(E)=\int\limits_{D_{i-1}}^{D_i}\tens{\mathcal{T}}_i\herm(z,E)
\tens{\sigma}^M_i(E)\tens{\mathcal{T}}_i(z,E) N_i W(z)\d z.\\
\end{split}
\label{eq:JcemsMulti}
\end{equation} 
Using the same approximations and omissions for \(\tens{\mathcal{T}}_i(z)\) as
 in deriving eq.~(\ref{eq:JcemsSimple}) from eq.~(\ref{eq:Jcems})  we may get
\begin{equation}
\begin{split}
\tens{J}_i(E)
=&-\sin\vartheta \int_{D_{i-1}}^{D_{i}} W(z) \e^{- N_i\sigma_i^e
\textstyle\frac{z-D_{i-1}}{\sin\vartheta}}\\
&\times\frac{\partial}{\partial z}\left(\e^{- N_i\tens{\sigma}_i^M(E)
\textstyle\frac{z-D_{i-1}}{\sin\vartheta}}\right)\d z. 
\end{split}\label{eq:JcemsSimpleMulti}
\end{equation}
Applying these expressions and knowing the proper \(W(z)\) escape function for the
 multilayer system the spectra can be calculated (provided that the approximations
 leading to (\ref{eq:JcemsSimple}) hold).

\section{\label{app:Transf} Transfer matrix} 
The form of eq.~(\ref{eq:Tdefinition}) is not readily available in \Refonlinecite{DeakPRB07}.
Therefore derivation of eq.~(\ref{eq:Tdefinition}) from formulae of \Refonlinecite{DeakPRB07}
is presented below.
 According the equations
 (A4-A5) of \Refonlinecite{DeakPRB07} we have
\begin{equation}
\tens{T}(z)=\tens{L}^{[22]}(\tens{I}+\tens{R})+
\tens{L}^{[21]}(\tens{I}-\tens{R}),\label{eq:TOffspAlak}
\end{equation}
 where the \(2\times2\) minor matrices of \(\tens{L}\) are defined in (B2)
 of \Refonlinecite{DeakPRB07}, as
\begin{equation}
\tens{L}^{[21]}=\imunit\sin\vartheta\tens{F}^{-1}\sinh\left(k_0 z \tens{F}\right),\quad
\tens{L}^{[22]}=\cosh\left(k_0 z \tens{F} \right), 
\end{equation}
where \(\tens{F}=\sqrt{-\tens{I}\sin^2\vartheta-\tens{\chi}}\). Using the
 definition (\ref{eq:mdef})
 we may write \(\tens{F}=\imunit\tens{m}\sin\vartheta\), and therefore
\begin{equation}
\tens{L}^{[21]}=\tens{m}^{-1}\sinh\left(\imunit k_0 z \tens{m}\sin\vartheta\right),\quad
\tens{L}^{[22]}=\cosh\left(\imunit k_0 z \tens{m}\sin\vartheta\right).
\end{equation}
Substituting these minor matrices of \(\tens{L}\) into (\ref{eq:TOffspAlak}) we
 get (\ref{eq:Tdefinition}).

\section{\label{app:Lilj} Electron escape function}

The electrons are emitted as a result of several secondary processes after irradiation the sample by
 \(\gamma\) photons, which all should be taken into account in  the escape functions.
According to \Refonlinecite{Liljequist78} the electron escape probability from a
 depth \(z\) can be written as:
\begin{equation}
W(z)=\sum\limits_i c_i W_i(z),
\end{equation}
where \(c_i\) gives the contribution of the \(i\)-th process, for which the escape 
function is \(W_i(z)\). The escape function, using the concept of the equivalent iron
 depth \(t\) instead of \(z\) (definition is given in (\ref{eq:tDefinition}) ) for K, L-M
 and Auger electrons, has the form:
\begin{equation}
W_i\left(x\right)=\left\{\begin{array}{ccc} 0.74-2.7x+2.5x^2& \text{if}&0\leqslant x \leqslant 0.55 \\
0&&\text{otherwise}\end{array}\right.,\label{eq:KLMAugerParabolic}
\end{equation}
where \(x=\frac{t}{r_i^\text{Bethe}}\) and \(r_i^\text{Bethe}\) is the Bethe
 range (which is on average the total distance, which an electron travels until it gets trapped)\cite{Liljequist78}.
Alternative formulae for K, L-M and Auger electrons based on Gaussian functions
 are available in \Refsonlinecite{Liljequist98}{Liljequist01}:
\begin{equation}
W(t,E_e)=A(Z,E_e)\e^{\textstyle-\frac{t}{R(Z,E_e)}}
\e^{\textstyle-\left(\frac{t}{1.9 R(Z,E_e)}\right)^2}, \label{eq:KLMAugerGauss}
\end{equation}
where \(Z\) is the atomic number and \(E_e\) the energy with which the electron
 was emitted from the atom. For further details regarding the functions \(A(Z,E_e)\)
 and \(R(Z,E_e)\) see \Refonlinecite{Liljequist98}. To get the alternative \(W_i(t)\)
 functions of (\ref{eq:KLMAugerParabolic}) we have to integrate \(W(t,E_e)\)
 according to the electron energies in the corresponding ranges too. Therefore
 (\ref{eq:KLMAugerGauss}) is more appropriate for DCEMS problems, for
 which (\ref{eq:KLMAugerParabolic}) is completely inappropriate.

For photo-electrons the escape function has the form:
\begin{widetext}
\begin{equation}
W_i(t)=\frac{{U_\infty}_i}{r^\text{eff}_i}\left\{\begin{array}{ccc}
1-0.5\left[\Expint{2}(\mu_i t)-\Expint{2}(\mu_i(r_i^\text{eff}-t))\right]& \text{if}
 & 0\leqslant t\leqslant r_i^\text{eff}\\
0.5\left[\Expint{2}(\mu_i(t-r_i^\text{eff}))-\Expint{2}(\mu_i t)\right]& \text{if}
 & r_i^\text{eff}<t
\end{array}\right.,
\end{equation}
\end{widetext}
where \(\Expint{2}(x)\) is the second-order exponential integral function defined by
 \(\Expint{2}(x)=\int_1^\infty h^{-2}\e^{-xh}\d h=\int_0^1\e^{-\frac{x}{u}}\d u\)
 \cite{AbramovitzStegun}, \(r_i^\text{eff}\) is the so-called effective range, \(\mu_i\)
 is absorption coefficient for the conversion photons, \({U_\infty}_i\) is a normalization constant.
\begin{acknowledgments}
Kind advices and helpful comments by Prof.\ Hartmut Spiering,
 Universit\"{a}t Mainz and financial support by Hungarian Scientific Research Fund in
 OTKA K62272 and T047094, and by the NAP-VENEUS'08 project of the National Office
 for Research and Technology of Hungary are gratefully acknowledged.
\end{acknowledgments}
%

\end{document}